\documentclass[preprint,showpacs,preprintnumbers,amsmath,amssymb,aps,nofootinbib]{revtex4}

\usepackage{graphicx}% Include figure files
\usepackage{dcolumn}% Align table columns on decimal point
\usepackage{bm}% bold math

\begin{document}

\title{Structure functions in deep inelastic scattering from gauge/string duality beyond  single-hadron final states  }

\author{Jian-hua Gao}
\email{gaojh@sdu.edu.cn}
\affiliation{Shandong Provincial Key Laboratory of Optical Astronomy and Solar-Terrestrial Environment, School of Space Science and Physics, Shandong University at Weihai, Weihai 264209, China}
 \affiliation{Key Laboratory of Quark and Lepton Physics (MOE), Central China Normal University, Wuhan 430079, China}

\author{Zong-Gang Mou}
\email{ppxzm1@nottingham.ac.uk}
\affiliation{School of Physics and Astronomy, University Park, University of Nottingham,
Nottingham NG7 2RD, United Kingdom}
\date{\today}

\begin{abstract}
We study deep inelastic scattering at large 't Hooft coupling and finite $x$ from
gauge/string duality beyond single-hadron final  states, which gives the leading large-$N_c$ contribution.
Within the supergravity approximation, we
calculate the  subleading large-$N_c$ contribution  by introducing an extra hadron into
the final states. We find that the  contribution from these double-hadron final states  will dominate in the Bjorken limit
$q^2\rightarrow \infty$ compared with the single-hadron states. We discuss the implications of our results.
\end{abstract}


\maketitle

\section{Introduction}
The gauge/string correspondence, since first conjectured \cite{Maldacena:1997re,Witten:1998qj,Gubser:1998bc},
has been used widely in studying nonperturbative aspects of
QCD. As the first application to deep inelastic scattering (DIS),  Polchinski and Strassler
\cite{Polchinski:2002jw}  employed the correspondence to calculate the structure functions for hadrons  in the large-'t Hooft coupling and large-$N_c$ limits by introducing an infrared cutoff in the fifth dimension  to mimic the confinement.
Since then, there have been many further investigations \cite{Brower:2007xg, Brower:2006ea,
BallonBayona:2008zi, BallonBayona:2007rs, BallonBayona:2007qr,
Pire:2008zf, Cornalba:2008sp,Albacete:2008ze, Levin:2009vj, Kovchegov:2009yj,Gao:2009ze,Hatta:2009ra,Gao:2010qk,Betemps:2010ij,Cornalba:2010vk,Agozzino:2013zgy,Koile:2013hba,Costa:2013uia}
in this direction.
Compared to QCD, these studies
show that the  hadron structures in the strong-coupling limit bear very different features at finite $x$,
while sharing similar features at small $x$.
It turns out that the structure functions in the strong-coupling regime are all power suppressed at finite $x$,
implying that few of the partons gain a finite amount of longitudinal momentum from the target hadron, and that almost all the partons are squeezed
into small-$x$ region. However, such conclusions can be arrived at only in the large-$N_c$ limit from
the contribution of single-hadron final states.
It is valuable to investigate the structure functions
% at finite $x$ at strong coupling
beyond this limit, which is   the major concern of the current work.

It is a nontrivial task to obtain complete contributions to the structure functions at  subleading order in large $N_c$ from gauge/string duality.
For simplicity, we will  restrict ourselves to the supergravity approximation,
considering subleading contributions from the processes with
only two scalar hadrons involved in final states.
Through the specific calculation and power analysis, we find that
in the large-$N_c$ expansion ( compared with leading contribution)
the subleading contribution can be less  suppressed in the power expansion of $1/q$.
Thus the subleading contribution
will  dominate in the Bjorken limit $q\rightarrow \infty$, which implies that large-$N_c$ limit and the Bjorken limit do not commute with each other.

The paper is organized as follows. In
Sec.\ref{duality} we formulate the DIS on a scalar target in the gauge/string correspondence.
In Sec.\ref{amplitude} we  evaluate successively the transition amplitudes, the hadronic tensor, and  structure functions for the DIS process
under the   supergravity approximation. In Sec.\ref{power} we analyze the power dependence on $1/q$ for various channels and phase spaces and
extract the leading contribution for the structure functions in the Bjorken limit $q\rightarrow \infty$. In Sec. \ref{conclusion}
we discuss our results and give a summary.

\section{DIS from the gauge/string duality}
\label{duality}
In the one-photon exchange approximation for DIS,  the initial lepton interacts with the hadron target  by the exchange of a virtual photon and
the  hadron absorbs the photon and decays into the final states. The cross section is determined by the hadronic tensor $W^{\mu\nu}$ which is defined as
\begin{equation}
  W^{\mu\nu} = \sum_X (2\pi)^4 \delta\left(p+q-P_X\right)
  \langle H | J^\mu(0)|X\rangle \langle X|  J^\nu(0)] | H \rangle \, ,
  \label{dis3}
\end{equation}
where $J^\mu$ is the electromagnetic current, $q^\mu$ is the momentum of the virtual photon, $p^\mu$ denotes the momentum of the initial hadron $H$ and
$P_X$ denotes the total momentum of the final hadron states $X$.
For the spinless  or spin-averaged hadrons, the hadronic tensor can be decomposed into
\begin{eqnarray}
\label{W-2}
W^{\mu\nu}
&=&
F_1\left(x,q^2\right)\left(\eta^{\mu\nu}-\frac{q^\mu q^\nu}{q^2}\right)
+\frac{2x}{q^2}F_2\left(x,q^2\right) \left(p^\mu + \frac{q^\mu}{2x}\right)\left(p^\nu + \frac{q^\nu}{2x}\right).
\end{eqnarray}%
All the information for the hadron structure is encoded in the structure functions $F_1\left(x,q^2\right)$ and $F_2\left(x,q^2\right)$.

In the gauge/string duality, scalar hadrons  correspond to normalizable  supergravity modes of the
dilaton and the electromagnetic  current corresponds to a non-normalizabel mode of a Kaluza-Klein gauge field at the boundary of  $\textrm{AdS}_5$.
The mass gap of hadrons can be generated by breaking the conformal invariance through introducing a sharp cutoff $0\leq z \leq z_0\equiv1/\Lambda$.
The metric in $\textrm{AdS}_5$ space can be written as
\begin{equation}
ds^{2}=\frac{1}{z^{2}}(\eta _{\mu \nu }dy^{\mu }dy^{\nu }+dz^{2})
\end{equation}%
where $\eta _{\mu \nu }=\left( -,+,+,+\right) $ is the flat-space metric at the boundary. The initial/final dilaton wave
function  satisfies the Klein-Gorden equation  in $\textrm{AdS}_5$ and the corresponding normalizable solution with the boundary
condition $\Phi (y,z_0)=0$ is given by
\begin{equation}
\Phi (y,z)=c_{\kappa,n} e^{ip\cdot y} z^{2}J_{\kappa}(M_{\kappa,n}z)
\end{equation}%
where $\kappa=\Delta-2$ with $\Delta$ being the conformal dimension of the state, $M_{\kappa,n} z_0$ denotes the $n$th zero point of the Bessel function $J_{\kappa}$ and $c_{\kappa,n}$ is the normalization factor,
\begin{equation}
\label{c}
c_{\kappa,n}=\frac{\sqrt{2}}{z_0|J_{\kappa+1}(M_{\kappa,n} z_0)|}.
\end{equation}
In order to calculate the subleading large-$N_c$ contribution from the final multiple-hadron states, we  need the bulk-to-bulk propagator of dilatons in $A\textrm{d}S_{5}$ space, which is given by
\begin{equation}
\label{propagator-5d-1}
G(y,z;y^\prime,z^{\prime })=-\int \frac{d^{4}k}{(2\pi )^{4}}e^{-ik\cdot
(y-y^\prime)}\int_{0}^{\infty }d\omega \frac{\omega }{\omega ^{2}+k^{2}-i\epsilon }%
z^{2}J_{\kappa}(\omega z)z^{\prime 2}J_{\kappa}(\omega
z^{\prime }),
\end{equation}%
When considering the boundary condition, the accurate propagator
should take the discrete form
\begin{equation}
\label{propagator-5d}
G(y,z;y^\prime,z^{\prime })=-\int \frac{d^{4}k}{(2\pi )^{4}}e^{-ik\cdot
(y-y^\prime)}\sum_{M_{\kappa,n}} \frac{ M_{\kappa, n}  c_{\kappa,n}^2}{M_{\kappa,n} ^{2}+k^{2}-i\epsilon }%
z^{2}J_{\kappa}(M_{\kappa,n} z)z^{\prime 2}J_{\kappa}(M_{\kappa,n}
z^{\prime }),
\end{equation}%
The gauge field  corresponding  to the current satisfies the Maxwell equations in $\textrm{AdS}_5$ space,
and  the non-normalizable solution with the boundary condition $A_{\mu }(y,\infty )=n_{\mu }e^{iq\cdot y}$ (where
$n_{\mu }$ is the virtual photon polarization vector) and the Lorentz-like gauge fixing $\partial _{\mu }A^{\mu }+z\partial _{z}\left( {A_{z}}/{z}\right) =0$
is given by
\begin{eqnarray}
A_{\mu } &=&n_{\mu }e^{iq\cdot y}qz{K}_{1}(qz),\ \ \   A_{z} =in\cdot qe^{iq\cdot y}z{K}_{0}(qz),
\end{eqnarray}%
where ${K}_{1}$ and ${K}_{0}$ are both modified Bessel functions.

\begin{figure}[tbp]
\begin{center}
\includegraphics[width=6cm]{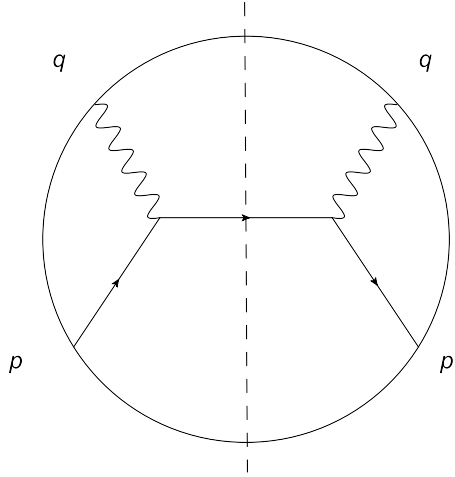}
\end{center}
\caption[]{Leading large-$N_c$ contribution from the single-hadron final states} \label{DIS1}
\end{figure}

When we were working in the leading large-$N_c$ approximation, only a single hadron in the final states was needed.
The corresponding  Witten diagram for the hadronic tensor is represented in Fig.\ref{DIS1}, in which the dashed
 line denotes the final states.  In our present work, we will devote ourselves to calculating the  subleading
large-$N_c$ contribution and  analyzing the power dependence of $1/q$. It should be mentioned that the complete subleading contribution of large
$N_c$ can come from different sources, however, in this paper we will only restrict ourselves to consider the contribution by introducing an extra hadron into the final states.
Doing  such a calculation is mainly inspired by the discussion of Polchinski and Strassler in \cite{Polchinski:2002jw}, while it should be emphasized  that it is possible that the ignored terms could cancel the leading $1/q$ contribution  or that they may have
even more important contributions in $1/q$ than those  found in the present  work. For simplicity, we will ignore these complexities in the following discussion.

For further simplicity, we will only consider the spinless hadron and the final states that include two dilatons, in which
the gauge propagator and gravity propagator do not contribute. The relevant supergravity interaction is
\begin{eqnarray}
\label{action}
S&=&-\int d^{5}x\sqrt{-g}\left[\sum_{i=1}^3 D^m \Phi_i D_m \Phi^{\ast }_i+\sum_{i=1}^3 \mu_i^2 \Phi_i^{\ast }\Phi_i
+\lambda\Phi_1\Phi_2^{\ast } \Phi_3^{\ast }+\lambda\Phi_1^{\ast }\Phi_2 \Phi_3\right]\nonumber\\
&=&-\int d^{5}x\sqrt{-g}\left[\sum_{i=1}^3\partial^m \Phi_i \partial_m \Phi^{\ast }_i+\sum_{i=1}^3 \mu_i^2 \Phi_i^{\ast }\Phi_i
+A^m A_m \sum_{i=1}^3{\cal Q}_i^2\Phi^{\ast }_i\Phi_i\right.\nonumber\\
& &\hspace{2.5cm}\left.+i A^m \sum_{i=1}^3{\cal Q}_i \left(\Phi_i\partial_m \Phi^{\ast }_i-\Phi^{\ast }_i\partial_m \Phi_i\right)
 +\lambda\Phi_1\Phi_2^{\ast } \Phi_3^{\ast }+\lambda\Phi_1^{\ast }\Phi_2 \Phi_3\right],
\end{eqnarray}%
where we have introduced three different dilatons ($i=1,2,3$) which have different charges ${\cal Q}_i$( with ${\cal Q}_1+{\cal Q}_2+{\cal Q}_3=0$) and five-dimensional mass $\mu_i^2=\Delta_i(\Delta_i-4)/R^2$ where $\Delta_i$ is the conformal dimension of the states and  $R$ is the AdS radius. It should be explained that the above action is mainly based on
 phenomenological considerations, and we assign the parameters ${\cal Q}_i$ and $\lambda$ as small, free coupling constants.

\begin{figure}[tbp]
\begin{center}
\includegraphics[width=6cm]{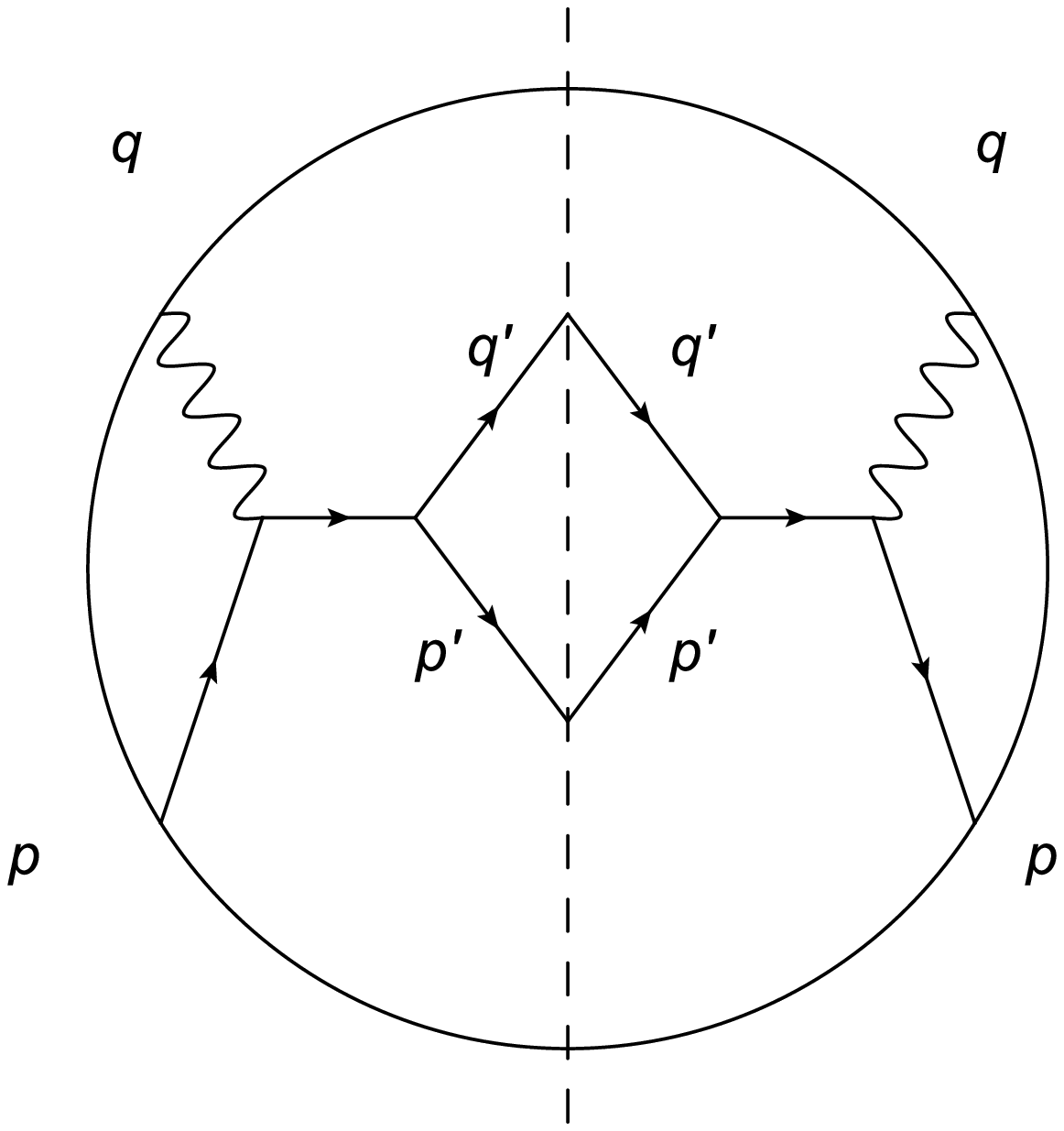}
\end{center}
\caption[*]{$s$-channel contribution from the double-hadron final states} \label{DIS2ss}
\end{figure}

\begin{figure}[tbp]
\begin{center}
\includegraphics[width=6cm]{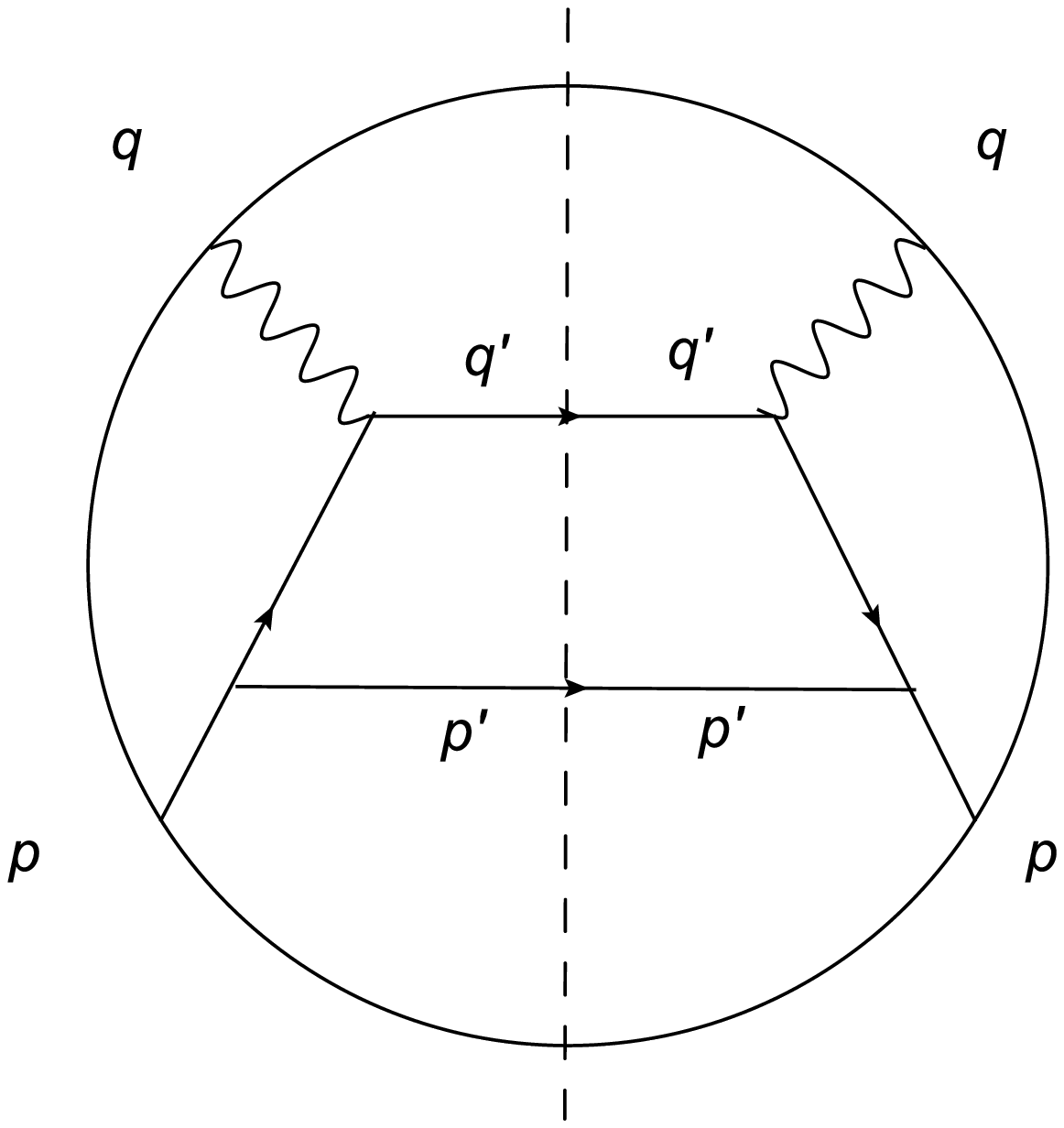}
\end{center}
\caption[*]{$t$-channel contribution from the double-hadron final states} \label{DIS2tt}
\end{figure}

\begin{figure}[tbp]
\begin{center}
\includegraphics[width=6cm]{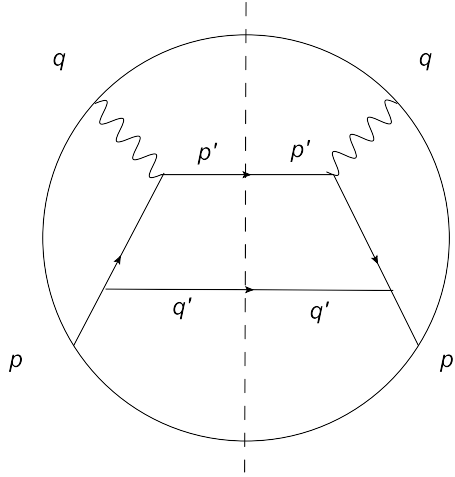}
\end{center}
\caption[*]{$u$-channel contribution from the double-hadron final states} \label{DIS2uu}
\end{figure}

It follows that the subleading contributions of large $N_c$ \footnote{In supergravity, the loop corrections are always equivalent to $1/N$-suppressed
corrections.  This  is because the action of interest to us [ e.g., Eq.(\ref{action})] will have an overall coupling-constant factor
which is proportional to $N^2$ (for brevity, we have suppressed this overall factor in this paper). Hence,  similar to the argument  in
large-$N_c$ QCD, we can find that the additional loops or external lines will result in an additional suppression of $1/N$. }
with two dilatons in the final states  come from the Witten diagrams
in Figs.\ref{DIS2ss},  \ref{DIS2tt}, and \ref{DIS2uu}, and from the other six crossed-channel diagrams which have not been displayed here. The Witten diagrams with a cutline
here actually represent the squared amplitudes: the  transition amplitude is to the left of the cut line, while the complex
conjugate is to the right. Therefore,in order to calculate the hadronic tensor we first need to calculate the transition amplitudes.  
From the  action given in Eq. (\ref{action}), it is straightforward to 
write down all the transition amplitudes corresponding to different channels: the \textit{s}-channel amplitude,
\begin{eqnarray}
\label{M-s}
{\cal M}_s
&=&i{\cal Q}_1\int d^{5}xd^{5}x^{\prime }\sqrt{-g(x)}\sqrt{-g(x^\prime)}
\Phi_{1}(x)A^M(x)\left[\partial_M G(x,x^{\prime })\right]\Phi_2^\ast (x^{\prime })\Phi _{3}^{\ast }(x^{\prime })\nonumber\\
& &-i{\cal Q}_1\int d^{5}xd^{5}x^{\prime }\sqrt{-g(x)}\sqrt{-g(x^\prime)}
\left[\partial_M\Phi_{1}(x)\right]A^M(x) G(x,x^{\prime })\Phi_2^\ast (x^{\prime })\Phi _{3}^{\ast }(x^{\prime })
\end{eqnarray}%
the \textit{t}-channel amplitude,
\begin{eqnarray}
\label{M-t}
{\cal M}_t
&=&-i{\cal Q}_2\int d^{5}xd^{5}x^{\prime }\sqrt{-g(x)}\sqrt{-g(x^\prime)}
\Phi_{1}(x)\Phi_3^\ast (x)\left[\partial_M^\prime G(x,x^{\prime })\right]A^M(x^\prime)\Phi _{2}^{\ast }(x^{\prime })\nonumber\\
& &+i{\cal Q}_2\int d^{5}xd^{5}x^{\prime }\sqrt{-g(x)}\sqrt{-g(x^\prime)}
\Phi_{1}(x)\Phi_3^\ast (x) G(x,x^{\prime })A^M(x^\prime)\left[\partial_M^\prime \Phi _{2}^{\ast }(x^{\prime })\right]
\end{eqnarray}%
and the \textit{u}-channel amplitude,
\begin{eqnarray}
\label{M-u}
{\cal M}_u
&=&-i{\cal Q}_3\int d^{5}xd^{5}x^{\prime }\sqrt{-g(x)}\sqrt{-g(x^\prime)}
\Phi_{1}(x)\Phi_2^\ast (x)\left[\partial_M^\prime G(x,x^{\prime })\right]A^M(x^\prime)\Phi _{3}^{\ast }(x^{\prime })\nonumber\\
& &+i{\cal Q}_3\int d^{5}xd^{5}x^{\prime }\sqrt{-g(x)}\sqrt{-g(x^\prime)}
\Phi_{1}(x)\Phi_2^\ast (x) G(x,x^{\prime })A^M(x^\prime)\left[\partial_M^\prime \Phi _{3}^{\ast }(x^{\prime })\right]
\end{eqnarray}%
where $x=(y,z),x'=(y',x')$, and
\begin{eqnarray}
\label{states}
A^{\mu }(x)&=&n^{\mu } e^{iq\cdot y}qz^3{K}_{1}(qz),\quad
A^{z}(x) = i n\cdot q e^{iq\cdot y}z^3{K}_{0}(qz),\nonumber\\
\Phi _{1}(x)&=&c_{1}z^{2}
J_{\kappa_1}(M_1 z)e^{ip\cdot y},\quad
\Phi _{2}^{\ast }(x)
=c_{2}z^{2}J_{\kappa_2}(M_2 z)e^{-iq^{\prime }\cdot y},\nonumber\\
\Phi _{3}^\ast(x)
&=&c_{3}z^{2}J_{\kappa_3}(M_3 z)e^{-ip^\prime\cdot y}
\end{eqnarray}%
For brevity, we have used the shorthand $c_i, M_i (i=1,2,3)$   for $c_{(i){n,k}}$ and $M_{(i){n,k}}$, where $k$ labels the conformal weight and $n$ labels the state.

The main task of the remaining parts of this work is to calculate the above transition amplitudes,
 square them to obtain the hadronic tensor, and finally extract the structure functions.

\section{Calculation of the structure functions}
\label{amplitude}

Substituting the  wave functions of the initial or final states in Eq.(\ref{states}) into the transition amplitudes
and integrating out the boundary coordinates $y$ and $y^\prime$ yields
\begin{eqnarray}
\label{M-s-2}
{\cal M}_s
&=&\ {\cal Q}_1c_1c_2c_3 (2\pi)^4\delta^4\left(p+q-p^\prime-q^\prime\right) n\cdot\left(2p+\frac{1}{x}q\right)\nonumber\\
& &\times\int dz dz^\prime  \frac{q}{z^\prime} J_{\kappa_1 }(M_1 z)K_1(qz) G_s(z,z^\prime) J_{\kappa_2}(M_2 z^\prime)J_{\kappa_3}(M_3 z^\prime)\nonumber\\
& &-\ {\cal Q}_1 (2\pi)^4\delta^4\left(p+q-p^\prime-q^\prime\right) n\cdot q\nonumber\\
& &\times\frac{1}{q}\int dz^\prime z^{\prime 2}J_{\kappa_1}(M_1 z^\prime)K_1(q z^\prime)J_{\kappa_2}(M_2 z^\prime)J_{\kappa_3}(M_3 z^\prime)
\end{eqnarray}%
\begin{eqnarray}
\label{M-t-2}
{\cal M}_t
&=&\ {\cal Q}_2c_1c_2c_3 (2\pi)^4\delta^4\left(p+q-p^\prime-q^\prime\right) n\cdot(2q^\prime+\frac{1}{y^\prime}q)\nonumber\\
& &\times\int dz dz^\prime \frac{q}{z} J_{\kappa_1}(M_1 z)J_{\kappa_3}(M_3^\prime z) G_t(z,z^\prime) K_1(q z^\prime)J_{\kappa_2}(M_2 z^\prime)\nonumber\\
& &+\ {\cal Q}_2 (2\pi)^4\delta^4\left(p+q-p^\prime-q^\prime\right) n\cdot q\nonumber\\
& &\times\frac{1}{q}\int dz z^{2}J_{\kappa_1}(M_1 z)K_1(q z)J_{\kappa_2}(M_2 z)J_{\kappa_3}(M_3 z)
\end{eqnarray}%
\begin{eqnarray}
\label{M-u-2}
{\cal M}_u
&=&\ {\cal Q}_3c_1c_2c_3 (2\pi)^4\delta^4\left(p+q-p^\prime-q^\prime\right) n\cdot(2p^\prime+\frac{1}{x^\prime}q)\nonumber\\
& &\times\int dz dz^\prime \frac{q}{z} J_{\kappa_1}(M_1 z)J_{\kappa_2}(M_2^\prime z) G_u(z,z^\prime) K_1(q z^\prime)J_{\kappa_3}(M_3 z^\prime)\nonumber\\
& &+\ {\cal Q}_3 (2\pi)^4\delta^4\left(p+q-p^\prime-q^\prime\right) n\cdot q\nonumber\\
& &\times\frac{1}{q}\int dz z^{2}J_{\kappa_1}(M_1 z)K_1(q z)J_{\kappa_2}(M_2 z)J_{\kappa_3}(M_3 z)
\end{eqnarray}%
where we have defined three scalar variables,
\begin{eqnarray}
x=-\frac{q^2}{2p\cdot q}\,,\ \ \ x^\prime=-\frac{q^2}{2p^\prime\cdot q}\,,\ \ \ y^\prime=-\frac{q^2}{2q^\prime\cdot q}
\end{eqnarray}%
and the reduced bulk-to-bulk propagators in the holographic  radial coordinate, which are given by
\begin{eqnarray}
\label{propagator-1d-s}
G_s(z,z^{\prime })&=&-\int_{0}^{\infty }d\omega \frac{ \omega c_s^2 }{\omega ^{2}+(p+q)^{2}-i\epsilon }%
z^{2}J_{\kappa_1}(\omega z)z^{\prime 2}J_{\kappa_1}(\omega
z^{\prime }),\\
\label{propagator-1d-t}
G_t(z,z^{\prime })&=&-\int_{0}^{\infty }d\omega \frac{\omega c_t^2  }{\omega ^{2}+(p^\prime-p)^{2}-i\epsilon }%
z^{2}J_{\kappa_2}(\omega z)z^{\prime 2}J_{\kappa_2}(\omega
z^{\prime }),\\
\label{propagator-1d-u}
G_u(z,z^{\prime })&=&-\int_{0}^{\infty }d\omega \frac{\omega c_u^2  }{\omega ^{2}+(p^\prime-q)^{2}-i\epsilon }%
z^{2}J_{\kappa_3}(\omega z)z^{\prime 2}J_{\kappa_3}(\omega
z^{\prime }),
\end{eqnarray}
which correspond to  the $s$-channel, $t$-channel, and $u$-channel, respectively. It should be noted that for brevity
we will use the integral notation instead of the sum notation  in the propagator. In order to be consistent with the cutoff in the AdS space,
we have introduced the normalization factors $c_s$, $c_t$, and $c_u$ which are given (respectively) by
\begin{eqnarray}
c_{s}&=&\frac{\sqrt{2}}{z_0\left(|J_{\kappa_1+1}(\omega z_0)|+|J_{\kappa_1}(\omega z_0)|\right)},\\
c_{t}&=&\frac{\sqrt{2}}{z_0\left(|J_{\kappa_2+1}(\omega z_0)|+|J_{\kappa_2}(\omega z_0)|\right)}\\
c_{u}&=&\frac{\sqrt{2}}{z_0\left(|J_{\kappa_3+1}(\omega z_0)|+|J_{\kappa_3}(\omega z_0)|\right)}.
\end{eqnarray}
The above normalization is very proper because it is always finite and will reduce to the usual normalization (\ref{c})
when the propagator is on-shell.

The total transition amplitude is obtained by summing over all the contributions from different channels,
\begin{eqnarray}
\label{M-total}
{\cal M}&=&{\cal M}_s+{\cal M}_u+{\cal M}_t\nonumber\\
&=& c_1 c_2 c_3 (2\pi)^4\delta^4\left(p+q-p^\prime-q^\prime\right)\nonumber\\
& &\times\left[
 n\cdot\left(2p+\frac{1}{x}q\right)C_s + n\cdot\left(2q^\prime+\frac{1}{y^\prime}q\right)C_t
 + n\cdot\left(2p^\prime+\frac{1}{x^\prime}q\right)C_u\right],
\end{eqnarray}%
where we have defined
\begin{eqnarray}
\label{C-s}
C_s&=&
{\cal Q}_1\int dz dz^\prime  \frac{q}{z^\prime} J_{\kappa_1}(M_1 z)K_1(qz) G_s(z,z^\prime) J_{\kappa_2}(M_2 z^\prime)J_{\kappa_3}(M_3 z^\prime),\\
\label{C-t}
C_t&=&
{\cal Q}_2\int dz dz^\prime \frac{q}{z} J_{\kappa_1}(M_1 z)J_{\kappa_3}(M_3 z) G_t(z,z^\prime) K_1(q z^\prime)J_{\kappa_2}(M_2 z^\prime),\\
\label{C-u}
C_u&=&
{\cal Q}_3\int dz dz^\prime \frac{q}{z} J_{\kappa_1}(M_1 z)J_{\kappa_2}(M_2 z) G_u(z,z^\prime) K_1(q z^\prime)J_{\kappa_3}(M_3 z^\prime).
\end{eqnarray}%
From the relation between the hadronic tensor and the squared transition amplitude,
\begin{eqnarray}
n^\mu n^\nu W_{\mu\nu}={\cal M}{\cal M}^*,
\end{eqnarray}
and the definitions of the structure functions in Eq.(\ref{W-2}), we can extract
the structure functions in the Bjorken limit $q\rightarrow \infty$ with $x$ fixed,
\begin{eqnarray}
\label{F-1-b}
F_1\left(x,q^2\right)
&=&c_1^2\sum_{M_2}\sum_{M_3} c_2^2 c_3^2 \int \frac{d^3\bf{p^\prime}}{2E_{p^\prime}(2\pi)^3} \frac{d^3\bf{q^\prime}}{2E_{q^\prime}(2\pi)^3}
(2\pi)^4\delta^4\left(p+q-p^\prime-q^\prime\right)\nonumber\\
& &\times 2 q^2\left\{ \left[v_u^{2}+4x^2(v_s\cdot v_u)^2\right]C_u C_u^*
+\left[v_t^{2}+4x^2(v_s\cdot v_t)^2\right]C_t C_t^*\right.\nonumber\\
& &\left. \hspace{1cm}+\left[v_u\cdot v_t +4x^2(v_s\cdot v_u)(v_s\cdot v_t)\right]\left(C_u C_t^*+C_t C_u^*\right)\right\}\\
\label{F-2-b}
F_2\left(x,q^2\right)
&=&c_1^2\sum_{M_2}\sum_{M_3} c_2^2 c_3^2 \int \frac{d^3\bf{p^\prime}}{2E_{p^\prime}(2\pi)^3} \frac{d^3\bf{q^\prime}}{2E_{q^\prime}(2\pi)^3}
(2\pi)^4\delta^4\left(p+q-p^\prime-q^\prime\right)\nonumber\\
& &\times4 x q^2 \left\{\left[v_s^{2}+12x^2v_s^4\right]C_s C_s^*
+\left[v_u^{2}+12x^2(v_u\cdot v_s)^2\right]C_u C_u^*
\right.\nonumber\\
& &\hspace{1cm}+\left[v_t^{2}+12x^2(v_t\cdot v_s)^2\right]C_t C_t^*+\left[v_s\cdot v_t +12x^2(v_t\cdot v_s) v_s^2\right]\left(C_s C_t^*+C_t C_s^*\right)\nonumber\\
& &\hspace{1cm}+\left[v_s\cdot v_u +12x^2(v_u\cdot v_s) v_s^2\right]\left(C_s C_u^*+C_u C_s^*\right)\nonumber\\
& &\left.\hspace{1cm}+\left[v_u\cdot v_t +12x^2(v_t\cdot v_s)(v_u\cdot v_s)\right]\left(C_u C_t^*+C_t C_u^*\right)\right\}
\end{eqnarray}%
where we have defined three vectors
\begin{eqnarray}
v_s^\mu&=&\frac{1}{q}\left(p^\mu +\frac{q^\mu}{2x}\right),\ \
v_u^{\mu}=\frac{1}{q}\left(p^{\prime\mu} +\frac{q^\mu}{2x^\prime}\right),\
v_t^{\mu}=\frac{1}{q}\left(q^{\prime\mu} +\frac{q^\mu}{2y^\prime}\right)
\end{eqnarray}%
In order to extract the leading contribution  in the Bjorken limit $q\rightarrow \infty$ with $x$ fixed, it is convenient
to define the following scaled variables:
\begin{eqnarray}
 \hat p^{\prime\mu}&=&p^{\prime\mu}/q, \ \ \hat q^{\prime\mu}=q^{\prime\mu}/q, \ \ \hat\omega =\omega/q,
 \ \ \hat z=q z,\ \ \hat z^\prime =q z^\prime
\end{eqnarray}%
With these scaled variables, we can rewrite the structure functions as
\begin{eqnarray}
\label{F-1-b-1}
F_1\left(x,q^2\right)
&=&c_1^2\sum_{M_2}\sum_{M_3} c_2^2 c_3^2 \int \frac{d^3 \bf{\hat p^\prime}}{2\hat {E}_{p^\prime}(2\pi)^3} \frac{d^3\bf{\hat q^\prime}}{2\hat {E}_{q^\prime}(2\pi)^3}
(2\pi)^4\delta^4\left(\hat p+\hat q-\hat p^\prime-\hat q^\prime\right)\nonumber\\
& &\times\frac{ 2}{ q^6} \left\{ \left[v_u^{2}+4x^2(v_s\cdot v_u)^2\right]\hat C_u \hat C_u^*
+\left[v_t^{2}+4x^2(v_s\cdot v_t)^2\right]\hat C_t \hat C_t^*\right.\nonumber\\
& &\left. \hspace{1cm}+\left[v_u\cdot v_t +4x^2(v_s\cdot v_u)(v_s\cdot v_t)\right]\left(\hat C_u \hat C_t^*+\hat C_t \hat C_u^*\right)\right\}\\
\label{F-2-b-1}
F_2\left(x,q^2\right)
&=&c_1^2\sum_{M_2}\sum_{M_3} c_2^2 c_3^2 \int \frac{d^3\bf{\hat p^\prime}}{2\hat E_{p^\prime}(2\pi)^3} \frac{d^3\bf{\hat q^\prime}}{2\hat E_{q^\prime}(2\pi)^3}
(2\pi)^4\delta^4\left(\hat p+\hat q-\hat p^\prime-\hat q^\prime\right)\nonumber\\
& &\times \frac{4 x}{ q^6} \left\{\left[v_s^{2}+12x^2v_s^4\right]\hat C_s\hat  C_s^*
+\left[v_u^{2}+12x^2(v_u\cdot v_s)^2\right]\hat C_u \hat C_u^*
\right.\nonumber\\
& &\hspace{1cm}+\left[v_t^{2}+12x^2(v_t\cdot v_s)^2\right]\hat C_t\hat  C_t^*+\left[v_s\cdot v_t +12x^2(v_t\cdot v_s) v_s^2\right]
\left(\hat C_s \hat C_t^*+\hat C_t\hat  C_s^*\right)\nonumber\\
& &\hspace{1cm}+\left[v_s\cdot v_u +12x^2(v_u\cdot v_s) v_s^2\right]\left(\hat C_s \hat C_u^*+\hat C_u\hat  C_s^*\right)\nonumber\\
& &\left.\hspace{1cm}+\left[v_u\cdot v_t +12x^2(v_t\cdot v_s)(v_u\cdot v_s)\right]\left(\hat C_u \hat C_t^*+\hat C_t \hat C_u^*\right)\right\}
\end{eqnarray}%
where
\begin{eqnarray}
\label{hat_C-s}
\hat C_s&=&
{\cal Q}_1\int d\hat z d\hat z^\prime  \frac{1}{\hat z^\prime} J_{\kappa_1}(\hat M_1 \hat z)K_1(\hat z) \hat G_s(\hat z,\hat z^\prime)
J_{\kappa_2}(\hat M_2 \hat z^\prime)J_{\kappa_3}(\hat M_3 \hat z^\prime)\\
\label{hat C-u}
\hat C_u&=&
{\cal Q}_3\int d\hat z d\hat z^\prime \frac{1}{\hat z} J_{\kappa_1}(\hat M_1 \hat z)J_{\kappa_2}(\hat M_2 \hat z) \hat G_u(\hat z,\hat z^\prime)
K_1(\hat  z^\prime)J_{\kappa_3 }(\hat M_3 \hat z^\prime)\\
\label{hat C-u}
\hat C_t&=&
{\cal Q}_2\int d\hat z d\hat z^\prime \frac{1}{\hat z} J_{\kappa_1}(\hat M_1 \hat z)J_{\kappa_3}(\hat M_3 \hat z) \hat G_t(\hat z,\hat z^\prime)
K_1(\hat  z^\prime)J_{\kappa_2 }(\hat M_2 \hat z^\prime)
\end{eqnarray}%
with
\begin{eqnarray}
\label{hatpropagator-1d-s}
\hat G_s(\hat z,\hat z^{\prime })&=&-\int_{0}^{\infty }d\hat \omega \frac{\hat \omega c_s^2 }{\hat \omega ^{2}-\hat s -i\epsilon }%
\hat z^{2}J_{\kappa_1}(\hat \omega\hat  z)\hat z^{\prime 2}J_{\kappa_1}(\hat \omega
\hat z^{\prime }),\\
\label{hatpropagator-1d-t}
\hat G_t(\hat z,\hat z^{\prime })&=&-\int_{0}^{\infty }d\hat \omega \frac{\hat \omega c_t^2}{\hat \omega ^{2}-\hat t -i\epsilon }%
\hat z^{2}J_{\kappa_2}(\hat \omega \hat z)\hat z^{\prime 2}J_{\kappa_2}(\hat \omega \hat z^{\prime }),\\
\label{hatpropagator-1d-u}
\hat G_u(\hat z,\hat z^{\prime })&=&-\int_{0}^{\infty }d\hat \omega \frac{\hat \omega c_u^2 }{\hat \omega ^{2}-\hat u -i\epsilon }%
\hat z^{2}J_{\kappa_3}(\hat \omega \hat z)\hat z^{\prime 2}J_{\kappa_3}(\hat \omega \hat z^{\prime })
\end{eqnarray}
It can be noticed that we need to deal with the integrals over triple Bessel functions, which in general cannot be calculated analytically.
However, we can choose some special cases, e.g., we can set $\kappa_1=1$, $\kappa_2=0$, and $\kappa_3=1$ and use the integral formula \cite{Table},
\begin{eqnarray}
\int_0^\infty d \hat z  \hat z^{2} J_{0}( a \hat  z) J_{0}(b \hat z) K_1(\hat z)
&=&\frac{2\sqrt{2}(a^2+b^2+1)}{\left[(a+b)^2+1\right]^{\frac{3}{2}}\left[(a-b)^2+1\right]^{\frac{3}{2}}},\\
\int_0^\infty d \hat z  \hat z^{2} J_{1}( a \hat  z) J_{1}(b \hat z) K_1(\hat z)
&=&\frac{8\sqrt{2} ab }{\left[(a+b)^2+1\right]^{\frac{3}{2}}\left[(a-b)^2+1\right]^{\frac{3}{2}}},\\
\int_0^\infty d \hat z  \hat z J_{1}( a \hat  z) J_{1}(b \hat z) J_0(c \hat z)
&=&\frac{\sqrt{2}(a^2+b^2-c^2) \Theta(a+b-c)\Theta(c-|a-b|)}{\pi^2 ab \left[(a+b)^2-c^2\right]^{\frac{1}{2}}\left[c^2-(a-b)^2\right]^{\frac{1}{2}} },
\end{eqnarray}
We then have
\begin{eqnarray}
\label{hat_C-s1}
\hat C_s
&=&-{\cal Q}_1\int_{|\hat M_3-\hat M_2|}^{\hat M_3+\hat M_2 }d\hat \omega \frac{\hat \omega c_s^2}{\hat \omega ^{2}-\hat s -i\epsilon }
\frac{8\sqrt{2}\hat M_1 \hat \omega }{\left[(\hat M_1 + \hat\omega)^2+1\right]^{\frac{3}{2}}\left[(\hat M_1 -\hat \omega)^2+1\right]^{\frac{3}{2}}}\nonumber\\
& &\hspace{2cm}\times\frac{\sqrt{2}\left[\hat M_3^2+\hat \omega^2 -\hat M_2^2\right]}{\hat M_3 \hat\omega \left[(\hat M_3 +\hat M_2 )^2-\hat\omega^2\right]^{\frac{1}{2}}
\left[\hat\omega^2-(\hat M_3 - \hat M_2)^2\right]^{\frac{1}{2}}}\hspace{2.5cm}
\end{eqnarray}%
\begin{eqnarray}
\label{hat C-t1}
\hat C_t
&=&-{\cal Q}_2\int_{|\hat M_3 - \hat M_1|}^{\hat M_3 + \hat M_1 }d\hat \omega \frac{\hat \omega c_t^2}{\hat \omega ^{2}-\hat t -i\epsilon }
\frac{\sqrt{2}\left[\hat M_1^2+ \hat M_3^2-\hat \omega^2\right]}
{\hat M_1 \hat M_3 \left[(\hat M_1 +\hat M_3)^2-\hat \omega^2\right]^{\frac{1}{2}}
\left|\hat \omega^2-(\hat M_1 - \hat M_3)^2\right|^{\frac{1}{2}}}\nonumber\\
& &\hspace{2cm}\times\frac{2\sqrt{2}\left(\hat M^2_2+\hat \omega^2+1 \right)}
{\left[(\hat M_2 + \hat\omega)^2+1\right]^{\frac{3}{2}}\left[(\hat M_2 -\hat \omega)^2+1\right]^{\frac{3}{2}}}
\end{eqnarray}%
\begin{eqnarray}
\label{hat C-u1}
\hat C_u
&=&-{\cal Q}_3\int_{|\hat M_2 - \hat M_1|}^{\hat M_2 + \hat M_1 }d\hat \omega \frac{\hat \omega c_u^2}{\hat \omega ^{2}-\hat u -i\epsilon }
\frac{\sqrt{2}\left[\hat M_1^2+\hat \omega^2 -\hat M^2_2\right]}
{\hat M_1\hat\omega\left[(\hat M_2 +\hat M_1 )^2-\hat\omega^2 \right]^{\frac{1}{2}}
\left[\hat\omega^2-(\hat M_2 - \hat M_1)^2\right]^{\frac{1}{2}}}\nonumber\\
& &\hspace{2cm}\times\frac{8\sqrt{2}\hat M_3 \hat \omega }
{\left[(\hat M_3 + \hat\omega)^2+1\right]^{\frac{3}{2}}\left[(\hat M_3 -\hat \omega)^2+1\right]^{\frac{3}{2}}}
\end{eqnarray}%
Now let us choose the center-of-mass frame of the initial dilaton and virtual photon, where
\begin{eqnarray}
p^\mu=\left(\frac{q}{2\sqrt{x(1-x)}},\frac{q}{2\sqrt{x(1-x)}},0,0\right),q^\mu=\left(\frac{(1-2x)q}{2\sqrt{x(1-x)}},-\frac{q}{2\sqrt{x(1-x)}},0,0\right)
\end{eqnarray}%
It follows that
\begin{eqnarray}
\label{F-1-b-2}
F_1\left(x,q^2\right)
&=&\sum_{M_2}\sum_{M_3}\frac{c_1^2 c_2^2 c_3^2|\hat{\bf p}'|}{4\pi q^6}\sqrt{\frac{x}{1-x}} \int d\theta \sin\theta\nonumber\\
& &\times\left\{ \left[v_u^{2}+4x^2(v_s\cdot v_u)^2\right]\hat C_u \hat C_u^*
+\left[v_t^{2}+4x^2(v_s\cdot v_t)^2\right]\hat C_t \hat C_t^*\right.\nonumber\\
& &\hspace{1cm}\left.
+\left[v_u\cdot v_t +4x^2(v_s\cdot v_u)(v_s\cdot v_t)\right]\left(\hat C_u \hat C_t^*+\hat C_t \hat C_u^*\right)\right\}\\
\label{F-2-b-2}
F_2\left(x,q^2\right)
&=&\sum_{M_2}\sum_{M_3}\frac{c_1^2 c_2^2 c_3^2|\hat{\bf p}'|x}{2\pi q^6}\sqrt{\frac{x}{1-x}} \int d\theta \sin\theta\nonumber\\
& &\times \left\{\left[v_s^{2}+12x^2v_s^4\right]\hat C_s\hat  C_s^*
+\left[v_u^{2}+12x^2(v_u\cdot v_s)^2\right]\hat C_u \hat C_u^*
+\left[v_t^{2}+12x^2(v_t\cdot v_s)^2\right]\hat C_t\hat  C_t^*
\right.\nonumber\\
& &\hspace{1cm}+\left[v_u\cdot v_t +12x^2(v_t\cdot v_s)(v_u\cdot v_s)\right]\left(\hat C_u \hat C_t^*+\hat C_t \hat C_u^*\right)\nonumber\\
& &\hspace{1cm}+\left[v_s\cdot v_t +12x^2(v_t\cdot v_s) v_s^2\right]
\left(\hat C_s \hat C_t^*+\hat C_t\hat  C_s^*\right)\nonumber\\
& &\left.\hspace{1cm}+\left[v_s\cdot v_u +12x^2(v_u\cdot v_s) v_s^2\right]\left(\hat C_s \hat C_u^*+\hat C_u\hat  C_s^*\right)\right\}
\end{eqnarray}
where $|\hat{\bf p}'|$ is determined by
\begin{eqnarray}
\sqrt{\frac{1-x}{x}}=\sqrt{\hat{\bf p}'^2+\hat M_3^2}+\sqrt{\hat{\bf p}'^2+\hat M_2^2}.
\end{eqnarray}

\section{Power analysis}
\label{power}
In order to extract the leading contribution  in the Bjorken limit $q\rightarrow \infty$,
we need to  analyze the power dependence of the structure functions on $1/q$ in different kinetic ranges.
In this work, we will always assume $\hat M_1 \ll 1$ for the initial hadron. Hence, we can classify the kinetic ranges into four different parts according to
the masses of the final hadrons: $\hat M_2 \sim 1 $ \& $\hat M_3 \sim 1 $, $\hat M_2 \ll 1 $ \& $\hat M_3 \sim 1 $, $\hat M_2 \sim 1 $ \& $\hat M_3 \ll 1 $, and
$\hat M_2 \ll 1 $ \& $\hat M_3 \ll 1 $. Now let us deal with them one by one.

\subsection{ $\hat M_2 \sim 1 $ \& $\hat M_3 \sim 1 $ }
In this region, we can reduce the integrals in Eqs.(\ref{hat_C-s1}-\ref{hat C-u1}) to
\begin{eqnarray}
\label{hat_C-s2-a}
\hat C_s
&\approx&-{\cal Q}_1 c_s^2 \frac{16\hat M_1}{\hat M_3}\int_{|\hat M_3-\hat M_2|}^{\hat M_3+\hat M_2 }
d\hat \omega \frac{}{ }
\frac{\hat \omega  }{\left(\hat \omega ^{2}-\hat s -i\epsilon\right)\left(\hat \omega^2+1\right)^3}\nonumber\\
& &\hspace{2cm}\times\frac{\left[\hat M_3^2+\hat \omega^2 -\hat M_2^2\right]}{  \left[(\hat M_3 +\hat M_2 )^2-\hat\omega^2\right]^{\frac{1}{2}}
\left[\hat\omega^2-(\hat M_3 - \hat M_2)^2\right]^{\frac{1}{2}}}
\end{eqnarray}
\begin{eqnarray}
\label{hat C-t2-a}
\hat C_t
&\approx&-{\cal Q}_2c_t^2\frac{4\pi\hat M_1 \hat M_3\left(\hat M^2_2+\hat M_3^2+1 \right)}
{\left[(\hat M_2 + \hat M_3)^2+1\right]^{\frac{3}{2}}\left[(\hat M_2 -\hat M_3)^2+1\right]^{\frac{3}{2}}
\left(\hat M_3^2 -\hat t\right)^2}\\
\label{hat C-u2-a}
\hat C_u
&\approx&{\cal Q}_3c_u^2\frac{16 \pi\hat M_1 \hat M_3 \hat u }
{\left[(\hat M_3 + \hat M_2)^2+1\right]^{\frac{3}{2}}\left[(\hat M_3 -\hat M_2)^2+1\right]^{\frac{3}{2}}
\left(\hat M_2^2 -\hat u\right)^2}.
\end{eqnarray}%
where we have simply  extracted  the normalization factors $c_s^2$, $c^2_t$, and $c^2_u$ and assigned them  the value that will give a dominant contribution,
\begin{eqnarray}
\label{c-stu}
c_{s}&=&\frac{\sqrt{2}}{z_0\left(|J_{\kappa_1+1}(\sqrt{s} z_0)|+|J_{\kappa_1}(\sqrt{s} z_0)|\right)},\\
c_{t}&=&\frac{\sqrt{2}}{z_0\left(|J_{\kappa_2+1}(\sqrt{|t|} z_0)|+|J_{\kappa_2}(\sqrt{|t|} z_0)|\right)}\\
c_{u}&=&\frac{\sqrt{2}}{z_0\left(|J_{\kappa_3+1}(\sqrt{|u|} z_0)|+|J_{\kappa_3}(\sqrt{|u|} z_0)|\right)}
\end{eqnarray}
The normalization coefficient $c_1$ for the initial hadron in Eqs.(\ref{F-1-b-2}) and (\ref{F-2-b-2}) is always of order  unity when $\hat M_1 \ll 1$, while
the normalization coefficients $c_2\sim  q^{\frac{1}{2}},\ \ c_3\sim  q^{\frac{1}{2}}$ for the final hadron states by using the asymptotic behavior of
the Bessel function. Besides, the sum over $M_2$ and $M_3$ contribute $\sum_{M_2}\sum_{M_3}\sim 1$. In order to obtain the final power behavior, we need to
divide  this kinetic region further according to the momenta of the internal propagators, which are given in Table I. A detailed analysis can be found
in Appendix B.
\begin{table}[h]
\label{table1}
\begin{tabular}{|c|c|c|c|c|}
\hline
\begin{tabular}{c}
kinetic region
\end{tabular}
 &
\begin{tabular}{c}
$c_s$,$c_u$,$c_t$
\end{tabular}
 &
\begin{tabular}{c}
$\hat C_s$,$\hat C_u$,$\hat C_t$
\end{tabular}
 &
\begin{tabular}{c}
phase space
\end{tabular}
 &
\begin{tabular}{c}
structure functions
\end{tabular} \\
\hline
\begin{tabular}{c}
$|\hat t| \sim 1$ \\ $|\hat u|\sim 1$
\end{tabular}
 &
\begin{tabular}{c}
$c_s\sim q^{\frac{1}{2}}$ \\
$c_u\sim q^{\frac{1}{2}}$ \\
$c_t\sim q^{\frac{1}{2}}$
\end{tabular}
 &
\begin{tabular}{c}
$\hat C_s \sim 1$ \\
$\hat C_u \sim 1$ \\
$\hat C_t \sim 1$
\end{tabular}
 &
 \begin{tabular}{c}
$\int \sin\theta d\theta\sim 1$
\end{tabular}
 &
\begin{tabular}{c}
$F_{ss}\sim \frac{1}{q^4}$, $F_{uu}\sim \frac{1}{q^4}$ \\
$F_{tt}\sim \frac{1}{q^4}$, $F_{su}\sim \frac{1}{q^4}$ \\
$F_{st}\sim \frac{1}{q^4}$, $F_{ut}\sim \frac{1}{q^4}$
 \end{tabular} \\
\hline
\begin{tabular}{c}
$|\hat t| \ll 1$
\\
$|\hat u| \sim 1$
\end{tabular}
 &
\begin{tabular}{c}
$c_s\sim q^{\frac{1}{2}} $
\\
$c_u\sim q^{\frac{1}{2}} $
\\
$c_t\sim 1 $
\end{tabular}
 &
\begin{tabular}{c}
$\hat C_s\sim 1$
\\
$\hat C_u\sim 1$
\\
$\hat C_t\sim \frac{1}{q}$
\end{tabular}
 &
  \begin{tabular}{c}
$\int\sin\theta d\theta\sim \frac{1}{q}$
\end{tabular}
 &
\begin{tabular}{c}
$F_{ss}\sim \frac{1}{q^5}$, $F_{uu}\sim \frac{1}{q^5}$ \\
$F_{tt}\sim \frac{1}{q^7}$, $F_{su}\sim \frac{1}{q^5}$ \\
 $F_{st}\sim \frac{1}{q^6}$, $F_{ut}\sim \frac{1}{q^6}$
\end{tabular}
\\
\hline
\begin{tabular}{c}
$|\hat t| \sim 1$
\\
$|\hat u| \ll 1$
\end{tabular}
 &
\begin{tabular}{c}
$c_s\sim q^{\frac{1}{2}} $
\\
$c_u\sim 1 $
\\
$c_t\sim q^{\frac{1}{2}} $
\end{tabular}
 &
\begin{tabular}{c}
$\hat C_s\sim 1$
\\
$\hat C_u\sim \frac{1}{q^2}$
\\
$\hat C_t\sim 1$
\end{tabular}
 &
  \begin{tabular}{c}
$\int \sin\theta d\theta\sim \frac{1}{q}$
\end{tabular}
 &
\begin{tabular}{c}
$F_{ss}\sim \frac{1}{q^5}$, $F_{uu}\sim \frac{1}{q^9}$\\
$F_{tt}\sim \frac{1}{q^5}$, $F_{su}\sim \frac{1}{q^7}$\\
$F_{st}\sim \frac{1}{q^6}$, $F_{ut}\sim \frac{1}{q^7}$
\end{tabular}
 \\
\hline
\end{tabular}
\caption{$\hat M_2 \sim 1 $ \& $\hat M_3 \sim 1 $. }
\end{table}
It should be pointed out that the subscripts in the structure functions denote the contribution from different channels, e.g.,
$F_{ss}$ means the contribution to the structure functions
from the term $\hat C_s \hat C_s^*$,  $F_{ut}$ means the contribution  from $\hat C_u \hat C_t^*+\hat C_t \hat C_u^*$, and so on.
Since the power dependencies for $F_1$ and $F_2$ are the same, we have suppressed the subscript 1 or 2. When there are no corresponding
terms, e.g., $F_{ss}$, $F_{st}$, and $F_{su}$ in $F_1$, we will only denote the contribution in $F_2$.
It is easy to show that $\hat t +\hat u \sim 1$, which implies that we do not need to consider the case where $t\ll 1$ and $u\ll 1$ at the same time.
\subsection{$\hat M_2 \ll 1 $ \& $\hat M_3 \sim 1 $ }
In this region, we can have
\begin{eqnarray}
\label{hat_C-s2-b}
\hat C_s
&\approx& {\cal Q}_1 c_s^2\frac{16\pi\hat M_1\hat M_3\hat s}{\left(\hat M_3^2+1\right)^3\left(\hat s -\hat M_3^2\right)^2}
\end{eqnarray}
\begin{eqnarray}
\label{hat C-u2-b}
\hat C_u
&\approx&-{\cal Q}_3c_u^2 \frac{8\pi\hat M_3  }{\hat M_1 \left(\hat M_3^2+1\right)^3}
\frac{\hat M_1^2 -\hat M_2^2 +u + \sqrt{\left(\hat M_1^2 -\hat M_2^2 +u\right)^2-4u\hat M_1^2}}
{\sqrt{\left(\hat M_1^2 -\hat M_2^2 +u\right)^2-4u\hat M_1^2}}
\end{eqnarray}
\begin{eqnarray}
\label{hat C-t2-b}
\hat C_t
&\approx&-{\cal Q}_2c_t^2 \frac{4\pi\hat M_1\hat M_3  }{\left(\hat M_3^2+1\right)^2\left(\hat M_3^2 -t\right)^2}
\end{eqnarray}
The normalization coefficients $c_2\sim  1, c_3\sim  q^{\frac{1}{2}}$ and
the sum over $M_2$ and $M_3$ contribute $\sum_{M_2}\sum_{M_3}\sim q$. A detailed power analysis in different kinetic intervals
 is given in Table II.
\begin{table}[h]
\begin{tabular}{|c|c|c|c|c|}
\hline
\begin{tabular}{c}
kinetic region
\end{tabular}
 &
\begin{tabular}{c}
$c_s$,$c_u$,$c_t$
\end{tabular}
 &
\begin{tabular}{c}
$\hat C_s$,$\hat C_u$,$\hat C_t$
\end{tabular}
 &
\begin{tabular}{c}
phase space
\end{tabular}
 &
\begin{tabular}{c}
structure functions
\end{tabular} \\
\hline
\begin{tabular}{c}
$|\hat t| \sim 1$ \\ $|\hat u| \sim 1$
\end{tabular}
 &
\begin{tabular}{c}
$c_s\sim q^{\frac{1}{2}} $\\
$c_u\sim q^{\frac{1}{2}} $\\
$c_t\sim q^{\frac{1}{2}} $
\end{tabular}
 &
\begin{tabular}{c}
$\hat C_s \sim 1$\\
$\hat C_u\sim 1$\\
$\hat C_t\sim 1$
\end{tabular}
 &
 \begin{tabular}{c}
$\int \sin\theta d\theta\sim 1$
\end{tabular}
 &
\begin{tabular}{c}
$F_{ss}\sim \frac{1}{q^4}$, $F_{uu}\sim \frac{1}{q^4}$\\
$F_{tt}\sim \frac{1}{q^4}$, $F_{su}\sim \frac{1}{q^4}$\\
$F_{st}\sim \frac{1}{q^4}$, $F_{ut}\sim \frac{1}{q^4}$
 \end{tabular} \\
\hline
\begin{tabular}{c}
$|\hat t| \ll 1$\\
$|\hat u| \sim 1$
\end{tabular}
 &
\begin{tabular}{c}
$c_s\sim q^{\frac{1}{2}} $\\
$c_u\sim q^{\frac{1}{2}} $\\
$c_t\sim 1 $
\end{tabular}
 &
\begin{tabular}{c}
$\hat C_s \sim 1$\\
$\hat C_u\sim 1$\\
$\hat C_t\sim\frac{1}{q}$
\end{tabular}
 &
\begin{tabular}{c}
$\int \sin\theta d\theta\sim \frac{1}{q}$
\end{tabular}
 &
\begin{tabular}{c}
$F_{ss}\sim \frac{1}{q^5}$, $F_{uu}\sim \frac{1}{q^5}$\\
$F_{tt}\sim \frac{1}{q^7}$, $F_{su}\sim \frac{1}{q^5}$\\
$F_{st}\sim \frac{1}{q^6}$, $F_{ut}\sim \frac{1}{q^6}$
\end{tabular}
\\
\hline
\begin{tabular}{c}
$|\hat t| \sim 1$
\\
$|\hat u|\ll 1$
\end{tabular}
 &
\begin{tabular}{c}
$c_s\sim q^{\frac{1}{2}} $\\
$c_u\sim 1 $\\
$c_t\sim q^{\frac{1}{2}} $
\end{tabular}
 &
\begin{tabular}{c}
$\hat C_s \sim 1$\\
$\hat C_u \sim \frac{1}{q^2}$\\
$\hat C_t \sim 1$
\end{tabular}
 &
  \begin{tabular}{c}
$\int \sin\theta d\theta\sim \frac{1}{q}$
\end{tabular}
 &
\begin{tabular}{c}
$F_{ss}\sim \frac{1}{q^5}$, $F_{uu}\sim \frac{1}{q^9}$\\
$F_{tt}\sim \frac{1}{q^5}$, $F_{su}\sim \frac{1}{q^7}$\\
$F_{st}\sim \frac{1}{q^5}$, $F_{ut}\sim \frac{1}{q^7}$
\end{tabular}
 \\
\hline
\end{tabular}
\caption{$\hat M_2 \ll 1 $ \& $\hat M_3 \sim 1 $. }
\end{table}

\subsection{ $\hat M_2 \sim 1 $ \& $\hat M_3 \ll 1 $ }
In this region, we can have
\begin{eqnarray}
\label{hat_C-s2-c}
\hat C_s
&\approx&{\cal Q}_1 c_s^2\frac{16\pi\hat M_1\hat M_3\hat s}{\left(\hat M_2^2+1\right)^3\left(\hat s -\hat M_2^2\right)^2}\\
\label{hat C-u2-c}
\hat C_u
&\approx&{\cal Q}_3c_u^2 \frac{16 \pi\hat M_3 \hat M_1 u }{\left(\hat M_2^2+1\right)^3\left(\hat M_2^2 -u\right)^2}\\
\label{hat C-t2-c}
\hat C_t
&=&-{\cal Q}_2c_t^2 \frac{2\pi }{\hat M_1\hat M_3 \left(\hat M_2^2+1\right)^2}
\frac{\hat M_1^2 +\hat M_3^2 -t - \sqrt{\left(\hat M_1^2 +\hat M_3^2 -t\right)^2-4\hat M_1^2\hat M_3^2}}
{\sqrt{\left(\hat M_1^2 +\hat M_3^2 -t\right)^2-4\hat M_1^2\hat M_3^2}}
\end{eqnarray}
The normalization coefficients $c_2\sim  q^{\frac{1}{2}}, c_3\sim 1$ and
the sum over $M_2$ and $M_3$ contribute $\sum_{M_2}\sum_{M_3}\sim q$. A detailed power analysis in different kinetic intervals
 is given in Table III.
\begin{table}[h]
\begin{tabular}{|c|c|c|c|c|}
\hline
\begin{tabular}{c}
kinetic region
\end{tabular}
 &
\begin{tabular}{c}
$c_s$,$c_u$,$c_t$
\end{tabular}
 &
\begin{tabular}{c}
$\hat C_s$,$\hat C_u$,$\hat C_t$
\end{tabular}
&
\begin{tabular}{c}
phase space
\end{tabular}
 &
\begin{tabular}{c}
structure functions
\end{tabular}\\
\hline
\begin{tabular}{c}
$|\hat t|\sim 1$\\
$|\hat u| \sim 1$
\end{tabular}
 &
\begin{tabular}{c}
$c_s\sim q^{\frac{1}{2}} $\\
$c_u\sim q^{\frac{1}{2}} $\\
$c_t\sim q^{\frac{1}{2}} $
\end{tabular}
 &
\begin{tabular}{c}
$\hat C_s \sim \frac{1} {q}$ \\
$\hat C_u \sim \frac{1} {q}$ \\
$\hat C_t \sim \frac{1} {q}$
\end{tabular}
 &
 \begin{tabular}{c}
$\int \sin\theta d\theta\sim 1$
\end{tabular}
 &
\begin{tabular}{c}
$F_{ss}\sim \frac{1}{q^6}$, $F_{uu}\sim \frac{1}{q^6}$\\
$F_{tt}\sim \frac{1}{q^6}$, $F_{su}\sim \frac{1}{q^6}$\\
$F_{st}\sim \frac{1}{q^6}$, $F_{ut}\sim \frac{1}{q^6}$
\end{tabular}\\
\hline
\begin{tabular}{c}
$|\hat t| \ll 1$\\
$|\hat u| \sim 1$
\end{tabular}
 &
\begin{tabular}{c}
$c_s\sim q^{\frac{1}{2}} $\\
$c_u\sim q^{\frac{1}{2}} $\\
$c_t\sim 1 $
\end{tabular}
 &
\begin{tabular}{c}
$\hat C_s \sim \frac{1}{q}$\\
$\hat C_u \sim \frac{1}{q}$\\
$\hat C_t \sim {q^2}$
\end{tabular}
 &
 \begin{tabular}{c}
$\int \sin\theta d\theta\sim \frac{1}{q}$
\end{tabular}
 &
\begin{tabular}{c}
$F_{ss}\sim \frac{1}{q^8}$, $F_{uu}\sim \frac{1}{q^8}$\\
$F_{tt}\sim \frac{1}{q^2}$, $F_{su}\sim \frac{1}{q^8}$\\
$F_{st}\sim \frac{1}{q^5}$, $F_{ut}\sim \frac{1}{q^5}$
\end{tabular}
\\
\hline
\begin{tabular}{c}
$|t| \sim 1$\\
$|u | \ll 1$
\end{tabular}
 &
\begin{tabular}{c}
$c_s\sim q^{\frac{1}{2}} $\\
$c_u\sim 1 $\\
$c_t\sim q^{\frac{1}{2}} $
\end{tabular}
 &
\begin{tabular}{c}
$\hat C_s \sim  \frac{1}{q}$\\
$\hat C_u \sim \frac{1}{q^3}$\\
$\hat C_t \sim \frac{1}{q}$
\end{tabular}
 &
  \begin{tabular}{c}
$\int \sin\theta d\theta\sim \frac{1}{q}$
\end{tabular}
 &
\begin{tabular}{c}
$F_{ss}\sim \frac{1}{q^7}$, $F_{uu}\sim \frac{1}{q^{11}}$\\
$F_{tt}\sim \frac{1}{q^7}$, $F_{su}\sim \frac{1}{q^9}$\\
$F_{st}\sim \frac{1}{q^7}$, $F_{ut}\sim \frac{1}{q^9}$
\end{tabular}
\\
\hline
\end{tabular}
\caption{$\hat M_2 \sim 1 $ \& $\hat M_3 \ll 1 $. }
\end{table}

\subsection{ $\hat M_2 \ll 1 $ \& $\hat M_3 \ll 1 $ }
In this region, we can have
\begin{eqnarray}
\label{hat_C-s2-d}
\hat C_s
&\approx&{\cal Q}_1 c_s^2\frac{16\pi\hat M_1\hat M_3 }{\hat s},\\
\label{hat C-u2-d}
\hat C_u
&\approx&-{\cal Q}_3c_u^2 \frac{8\pi\hat M_3  }{\hat M_1 }
\frac{\hat M_1^2 -\hat M_2^2 +u + \sqrt{\left(\hat M_1^2 -\hat M_2^2 +u\right)^2-4u\hat M_1^2}}
{\sqrt{\left(\hat M_1^2 -\hat M_2^2 +u\right)^2-4u\hat M_1^2}},\\
\label{hat C-t2-d}
\hat C_t
&=&-{\cal Q}_2c_t^2 \frac{2\pi }{\hat M_1\hat M_3 }
\frac{\hat M_1^2 +\hat M_3^2 -t - \sqrt{\left(\hat M_1^2 +\hat M_3^2 -t\right)^2-4\hat M_1^2\hat M_3^2}}
{\sqrt{\left(\hat M_1^2 +\hat M_3^2 -t\right)^2-4\hat M_1^2\hat M_3^2}}.
\end{eqnarray}
The normalization coefficients  $c_2\sim  1 , c_3\sim 1$, and
the sum over $M_2$ and $M_3$ contribute $\sum_{M_2}\sum_{M_3}\sim 1$. The detailed power analysis in different kinetic intervals
 is given in Table IV.
\begin{table}[h]
\begin{tabular}{|c|c|c|c|c|}
\hline
\begin{tabular}{c}
kinetic region
\end{tabular}
 &
\begin{tabular}{c}
$c_s$,$c_u$,$c_t$
\end{tabular}
 &
\begin{tabular}{c}
$\hat C_s$,$\hat C_u$,$\hat C_t$
\end{tabular}
&
\begin{tabular}{c}
phase space
\end{tabular}
 &
\begin{tabular}{c}
structure functions
\end{tabular}\\
\hline
\begin{tabular}{c}
$|\hat t| \sim 1$\\
$|\hat u| \sim 1$
\end{tabular}
 &
\begin{tabular}{c}
$c_s\sim q^{\frac{1}{2}} $\\
$c_u\sim q^{\frac{1}{2}} $\\
$c_t\sim q^{\frac{1}{2}} $
\end{tabular}
 &
\begin{tabular}{c}
$\hat C_s \sim \frac{1} {q}$ \\
$\hat C_u \sim \frac{1} {q}$ \\
$\hat C_t \sim \frac{1} {q}$
\end{tabular}
 &
 \begin{tabular}{c}
$\int \sin\theta d\theta\sim 1$
\end{tabular}
 &
\begin{tabular}{c}
$F_{ss}\sim \frac{1}{q^8}$, $F_{uu}\sim \frac{1}{q^8}$\\
$F_{tt}\sim \frac{1}{q^8}$, $F_{su}\sim \frac{1}{q^8}$\\
$F_{st}\sim \frac{1}{q^8}$, $F_{ut}\sim \frac{1}{q^8}$
\end{tabular}\\
\hline
\begin{tabular}{c}
$|\hat t| \ll 1$\\
$|\hat u|\sim 1$
\end{tabular}
 &
\begin{tabular}{c}
$c_s\sim q^{\frac{1}{2}} $\\
$c_u\sim q^{\frac{1}{2}} $\\
$c_t\sim 1 $
\end{tabular}
 &
\begin{tabular}{c}
$\hat C_s \sim \frac{1}{q}$\\
$\hat C_u \sim \frac{1}{q}$\\
$\hat C_t \sim {q^2}$
\end{tabular}
 &
 \begin{tabular}{c}
$\int \sin\theta d\theta\sim \frac{1}{q}$
\end{tabular}
 &
\begin{tabular}{c}
$F_{ss}\sim \frac{1}{q^9}$, $F_{uu}\sim \frac{1}{q^9}$\\
$F_{tt}\sim \frac{1}{q^3}$, $F_{su}\sim \frac{1}{q^9}$\\
$F_{st}\sim \frac{1}{q^6}$, $F_{ut}\sim \frac{1}{q^6}$
\end{tabular}
\\
\hline
\begin{tabular}{c}
$|t |\sim 1$\\
$|u|\ll 1$
\end{tabular}
 &
\begin{tabular}{c}
$c_s\sim q^{\frac{1}{2}} $\\
$c_u\sim 1 $\\
$c_t\sim q^{\frac{1}{2}} $
\end{tabular}
 &
\begin{tabular}{c}
$\hat C_s \sim  \frac{1}{q}$\\
$\hat C_u \sim \frac{1}{q}$\\
$\hat C_t \sim \frac{1}{q}$
\end{tabular}
 &
  \begin{tabular}{c}
$\int \sin\theta d\theta\sim \frac{1}{q}$
\end{tabular}
 &
\begin{tabular}{c}
$F_{ss}\sim \frac{1}{q^9}$, $F_{uu}\sim \frac{1}{q^{9}}$\\
$F_{tt}\sim \frac{1}{q^9}$, $F_{su}\sim \frac{1}{q^9}$\\
$F_{st}\sim \frac{1}{q^9}$, $F_{ut}\sim \frac{1}{q^9}$
\end{tabular}
\\
\hline
\end{tabular}
\caption{$\hat M_2 \ll 1 $ \& $\hat M_3 \ll 1 $. }
\end{table}

\subsection{The final dominant contribution}

From the above analysis, we find that the dominant contribution is from the $t$-channel where $\hat M_1\ll 1$ $\hat M_3 \ll 1$, $\hat M_2\sim 1$
with $|\hat t| \ll 1$ and $|\hat u| \sim 1$. Hence, the leading contribution is given by
\begin{eqnarray}
\label{F-final}
F_1\left(x,q^2\right)
&\approx&\left(\frac{\Lambda}{q}\right)^2 f_1(x),\ \ \
F_2\left(x,q^2\right)
\approx\left(\frac{\Lambda}{q}\right)^2 f_2(x)
\end{eqnarray}
where we have extracted the power dependence and lumped all the others into the functions $f_1(x)$ and $f_2(x)$, which are independent of  $q$ or at
most dependent on $q$ by $\ln q $. Since  we are most interested in  the power dependence in this work, we will not present the
specific forms of $f_1(x)$ and $f_2(x)$.

\section{discussion and conclusion}
\label{conclusion}
Now let us compare the  results in Eq.(\ref{F-final}) from the subleading large-$N_c $ contribution with those obtained before from the
leading large-$N_c$ contribution, which is given by
\begin{eqnarray}
\label{F-b-2}
F_1\left(x,q^2\right)
&=&0\ \ \
F_2\left(x,q^2\right)
\approx\left(\frac{\Lambda}{q}\right)^{2\kappa_1 +2} f(x)
\end{eqnarray}
By setting $\kappa_1=1$ ( to be consistent with our present specific case), we have
\begin{eqnarray}
\label{F-b-2}
F_1\left(x,q^2\right)
&=&0\ \ \
F_2\left(x,q^2\right)
\approx\left(\frac{\Lambda}{q}\right)^{4} f(x)
\end{eqnarray}
We  notice two significant  differences between them. First, for the leading large-$N_c$ contribution,
the structure function $F_1\left(x,q^2\right)$ always vanishes, but it will obtain a nonzero contribution at  the subleading
large-$N_c$ order. As we all know, $F_1\left(x,q^2\right)$ is proportional to the Casimir of the scattered hadron under the
Lorentz transformation, so it is natural that $F_1\left(x,q^2\right)$ vanishes when the virtual photon hits the original
scalar target hadron directly at the leading large-$N_c$ order. However, at the subleading large-$N_c$ order, the scalar target hadron
can split into two scalar hadrons, and each hadron can have orbital angular momentum and  can lead to a nonvanishing $F_1\left(x,q^2\right)$
when they are hit by the virtual photon. Such arguments can be verified by Eq.(\ref{F-1-b}), in which only the $t$-channel and $u$-channel contribute
to $F_1\left(x,q^2\right)$, and the s-channel in which the target hadron interacts directly with the virtual photon does not contribute at all.
Second, the subleading large-$N_c$ contribution from the double-hadron final states is  less power-suppressed than the leading large-$N_c$ one.
The power dependence of the structure function is the same as that of  the hadron 2 from the leading large-$N_c$ contribution.
This conclusion makes sense, because in the dominant contribution (as discussed above)  the incoming hadron 1 splits into two hadrons 2 and 3; hadron 2 has the minimum twist
$\kappa_2=0$,  which propagates to the boundary of  AdS  $z=0$ and interacts with the current. When $\hat t \ll 1$, we can regard hadron 2 as an almost on-shell
hadron, and hence the final power dependence should be controlled by the twist of hadron 2. Our calculation and analysis verify this argument, which
was originally proposed in Ref. \cite{Polchinski:2002jw}. The result that the subleading contribution in  $N_c$ will dominate in the Bjorken limit
$q^2\rightarrow \infty$ implies that the large-$N_c$ limit and the Bjorken limit do not commute with each other.   Such a conclusion can lead to very important
consequences in  DIS from gauge/gravity duality. When we are calculating within supergravity, we {\it{a priori}} take the large $N_c$ limit first,  followed by
the Bjorken limit; however, when we are analyzing the process using operator product expansion, we  actually {\it{a priori}} take the Bjorken limit first,  followed by the
large-$N_c$ limit.   If  the  large-$N_c$ limit and the Bjorken limit do not commute with each other any more, this mutual comparison and
 analysis would lose valuable meaning. There is no doubt that we need  further investigation in this direction.
 We postpone such an investigation for a future work.

\begin{acknowledgments}
J.H.G.   was supported in part by  the Major State Basic Research Development Program in China (Grant No. 2014CB845406), 
the National Natural Science Foundation of China under  Grant No.~11105137, and  CCNU-QLPL Innovation Fund (QLPL2014P01).
\end{acknowledgments}

\begin{appendix}

\section{Derivation of the transition amplitude}
 In this appendix, we will derive the transition amplitudes in  Eqs.(\ref{M-s}-\ref{M-u}) from
 the  action given in   Eq.(\ref{action}).  As opposed to  the usual calculation of correlators for operators
 in the conformal field theory that lives at the boundary of the AdS space, here we are interested in the scattering process of physical
 states. We will follow the ansatz made by Polchinski and Strassler in \cite{Polchinski:2001tt}
 for the scattering of gauge-invariant states.
 The equations of motion corresponding to the action (\ref{action}) read
\begin{eqnarray}
\label{Phi1}
\frac{1}{\sqrt{-g}}\partial_m\left(g^{mn}\sqrt{-g}\partial_n \Phi_1\right)-\mu_1^2\Phi_1
&=& \frac{-i{\cal Q}_1}{\sqrt{-g}}  \partial_m \left(\sqrt{-g} A^m\Phi_1\right)- i{\cal Q}_1A^m\partial_m \Phi_1+\lambda \Phi_2^{\ast } \Phi_3^{\ast }\hspace{4pt}\\
\label{Phi2}
\frac{1}{\sqrt{-g}}\partial_m\left(g^{mn}\sqrt{-g}\partial_n \Phi_2\right)-\mu_2^2\Phi_2
&=& \frac{-i{\cal Q}_2}{\sqrt{-g}}  \partial_m \left(\sqrt{-g} A^m\Phi_2\right)- i{\cal Q}_2A^m\partial_m \Phi_2+\lambda \Phi_1^{\ast } \Phi_3\hspace{4pt}\\
\label{Phi3}
\frac{1}{\sqrt{-g}}\partial_m\left(g^{mn}\sqrt{-g}\partial_n \Phi_3\right)-\mu_3^2\Phi_3
&=& \frac{-i{\cal Q}_3}{\sqrt{-g}}  \partial_m \left(\sqrt{-g} A^m\Phi_3\right)- i{\cal Q}_3A^m\partial_m \Phi_3+\lambda \Phi_1^{\ast }\Phi_2
\end{eqnarray}
where we have suppressed  the terms $iQ_i^2 A^m A_m \Phi^{\ast}_i$ which are not relevant to the process we are considering.
The solution up to  first order in the coupling of ${\cal Q}_i$ or $\lambda$ is
\begin{eqnarray}
 \Phi_1(x)&=&
-i{\cal Q}_1 \int{d^5x'}G(x,x')
\partial_m \left[\sqrt{-g(x')} A^m(x')\Phi_1(x')\right]\nonumber\\
& &+ \int{d^5x'}\sqrt{-g(x')}G(x,x')
\left[-i{\cal Q}_1 A^m(x')\partial_m\Phi_1(x')+\lambda \Phi_2^{\ast }(x') \Phi_3^{\ast }(x')\frac{}{}\right]
\end{eqnarray}
Integrating the first term by parts gives
\begin{eqnarray}
 \Phi_1(x)&=&
i{\cal Q}_1 \int{d^5x'}\sqrt{-g(x')} \partial'_m G(x,x') A^m(x')\Phi_1(x')\nonumber\\
& &+ \int{d^5x'}\sqrt{-g(x')}G(x,x')
\left[-i{\cal Q}_1 A^m(x')\partial_m\Phi_1(x')+\lambda \Phi_2^{\ast }(x') \Phi_3^{\ast }(x')\frac{}{}\right]
\end{eqnarray}
The solution up to  second order in the coupling of ${\cal Q}_i$ or $\lambda$ is
\begin{eqnarray}
 \Phi_1(x)&=&
i{\cal Q}_1\lambda \int{d^5x'} d^5x''\sqrt{-g(x')} \sqrt{-g(x'')}\partial'_m G(x,x') A^m(x')G(x',x'')
\Phi_2^{\ast }(x'') \Phi_3^{\ast }(x'')\nonumber\\
& &-i{\cal Q}_1 \lambda \int{d^5x'}d^5x'' \sqrt{-g(x')}\sqrt{-g(x'')} G(x,x')
A^m(x')\partial'_m G(x',x'')\Phi_2^{\ast }(x'') \Phi_3^{\ast }(x'')\nonumber\\
& &-i{\cal Q}_2\lambda \int{d^5x'} d^5x''\sqrt{-g(x')} \sqrt{-g(x'')} G(x,x')\Phi_3^{\ast }(x')\partial'_m G(x',x'') A^m(x'')
\Phi_2^{\ast }(x'') \nonumber\\
& &+i{\cal Q}_2 \lambda \int{d^5x'}d^5x'' \sqrt{-g(x')}\sqrt{-g(x'')} G(x,x')\Phi_3^{\ast }(x')G(x',x'') A^m(x'')
\partial''_m \Phi_2^{\ast }(x'') \nonumber\\
& &-i{\cal Q}_3\lambda \int{d^5x'} d^5x''\sqrt{-g(x')} \sqrt{-g(x'')} G(x,x')\Phi_2^{\ast }(x')\partial'_m G(x',x'') A^m(x'')
\Phi_3^{\ast }(x'') \nonumber\\
& &+i{\cal Q}_3 \lambda \int{d^5x'}d^5x'' \sqrt{-g(x')}\sqrt{-g(x'')} G(x,x')\Phi_2^{\ast }(x')G(x',x'') A^m(x'')
\partial''_m \Phi_3^{\ast }(x'') \nonumber\\
\end{eqnarray}
It follows that the transition amplitudes (\ref{M-s}-\ref{M-u}) can be obtained by contracting the above expression with the
initial wave function in Eq.(\ref{states}). In order to obtain the final result, it should be noted that when $\Phi_1(x)$ is given
in Eq.(\ref{states}) the following identity holds:
\begin{eqnarray}
 \int{d^5x'} \sqrt{-g(x')}  G(x,x') \Phi_i(x') = \Phi_i(x).
\end{eqnarray}
The first and second terms correspond to the $s$-channel,
the third and fourth terms correspond to the $t$-channel, and the last two terms are the $u$-channel.
By using the formalism proposed by  Polchinski and Strssler in \cite{Polchinski:2001tt}
for the scattering process of gauge-invariant states, we do not need to deal with the boundary value of the fields.
Hence we do not meet any UV issues in our calculation. Besides, since what we are more interested in is  the power dependence
rather than the overall magnitude, UV issues are not very relevant in the present work.

\section{Detailed analysis of the power dependence}

In this appendix, we will take the first case where  $\hat M_2 \sim 1 $ and $\hat M_3 \sim 1 $ as an example and give
a detailed analysis of the power dependence. As we mentioned above, we will always assign $\hat M_1 \ll 1$ for the initial hadron.
Due to the kinematic constraint, the Mandelstam variable $\hat s$  is always of order  unity.
First, let us consider the power dependence from various of normalization factor:
$c_1$, $c_2$, $c_3$, $c_s$, $c_t$, and $c_u$. Recalling the definitions of these factors in Eqs.(\ref{c}) and (\ref{c-stu}), it is easy
to see that  $\hat M_1 \ll 1$, $|\hat t| \ll 1$, and $|\hat u| \ll 1$, i.e.,
$M_1 \ll q $, $|t|\ll q^2$, and $|u|\ll q^2$ yield $c_1\sim 1$, $c_t\sim 1$, and $c_u\sim 1$, respectively.
When we want to consider $\hat M_2 \sim 1$, $\hat M_3 \sim 1$,   $\hat s \sim 1$,  $|\hat t| \sim 1$, or $|\hat u| \sim 1$, i.e.,
$M_2 \sim q \rightarrow \infty $, $M_3 \sim q \rightarrow \infty $,  $s \sim q^2 \rightarrow \infty $,  $|t |\sim q^2 \rightarrow \infty $, or
$|u| \sim q^2 \rightarrow \infty $, we need  the asymptotic behavior of the Bessel function $J_\nu(z)$ in the limit $|z|\rightarrow \infty$,
\begin{eqnarray}
J_\nu(z)&\sim&\sqrt{\frac{2}{\pi z}}\cos\left(z-\frac{\nu \pi}{2}-\frac{\pi}{4}\right)
\end{eqnarray}
When $z\rightarrow \infty$, the $n$th zero point $z_n\sim (2n+1)\pi+\frac{\nu \pi}{2}+\frac{\pi}{4}\pm \frac{\pi}{2} $. With this
asymptotic expression, $\hat M_2 \sim 1$, $\hat M_3 \sim 1$,   $\hat s \sim 1$,  $|\hat t| \sim 1$ and  $|\hat u |\sim 1$ yield
$c_2\sim q^{\frac{1}{2}}$, $c_3\sim q^{\frac{1}{2}}$, $c_s\sim q^{\frac{1}{2}}$, $c_t\sim q^{\frac{1}{2}}$ and $c_u\sim q^{\frac{1}{2}}$ respectively.
From the expressions for $\hat C_s$ in Eq.(\ref{hat_C-s2-a}),
 $\hat s \sim 1$ yields  $\hat C_s \sim c_s^2 \hat M_1 \sim 1$.
From the expressions for $\hat C_t$  in Eq.(\ref{hat C-t2-a}), we  have $\hat C_t \sim c_t^2 \hat M_1 $. Then
 $|\hat t| \sim 1$ and  $|\hat t| \ll 1$ yield  $\hat C_t \sim 1 $ and $\hat C_t \sim  \frac{1}{q}$, respectively, and $|\hat t| \sim 1$ and  $|\hat u| \sim 1$ yield
 $\hat C_u \sim c_u^2 \hat M_1 \sim 1 $ and $\hat C_t \sim c_t^2 \hat M_1 \sim 1$, respectivle. 
In a similar way, from the expressions for $\hat C_u$  in Eq.(\ref{hat C-u2-a}) we  have $\hat C_u \sim c_u^2 \hat M_1 $. Then
 $|\hat u| \sim 1$ and  $|\hat u| \ll 1$ yield  $\hat C_u \sim 1 $ and $\hat C_u \sim  \frac{1}{q^2}$, respectively.
It is very straightforward ( but tedious) to verify that the factors involving $v_s^2$, $v_t^2$, $v_u^2$, $v_s\cdot v_u$, $v_u\cdot v_t$, and $v_s\cdot v_t$
in Eq.(\ref{F-1-b-2}) and (\ref{F-2-b-2}) can be of order unity. It is very easy to show that $|\hat{\bf p}'|$ can  always be of order  unity.
Hence, every term in $F_1\left(x,q^2\right)$ or $F_2\left(x,q^2\right)$ from different channels behaves as
\begin{eqnarray}
F_{\alpha\beta}\sim \frac{1}{q^6}c_1^2 c_2^2 c_3^2 \sum_{M_2}\sum_{M_3} \int d\theta \sin\theta
\left(\hat C_{\alpha }\hat C^\ast_{\beta}+\hat C_{\beta }\hat C^\ast_{\alpha}\right)
\end{eqnarray}
where both $\alpha$ and $\beta$  denote the different types of channels ($s$, $t$ and $u$). Since the hadrons with $\hat M_2\sim 1$ and $\hat M_3\sim 1$
are very limited, the sum over $M_2$ and $M_3$ contributes $\sum_{M_2}\sum_{M_3}\sim 1$.  The integral over the phase space
$\int d\theta \sin\theta$ depends on the interval of $\hat t$ or $\hat u$. $|\hat t| \sim 1$ and $|\hat u|\sim 1$ yield
$\int d\theta \sin\theta \sim 1$ and  $|\hat t| \sim 1$ and $|\hat u|\ll 1$ or $|\hat u| \sim 1$ and $|\hat t|\ll 1$ yield
$\int d\theta \sin\theta \sim \frac{1}{q}$. Putting all of these together, we can finally obtain the results in Table I.

\end{appendix}

\end{document}